\documentclass[aps,prd,showpacs,nofootinbib,floats,floatfix,preprintnumbers,groupedaddress,twocolumn]{revtex4}
\usepackage{bm}
\usepackage{latexsym}
\usepackage{dcolumn}
\usepackage{amsmath,amsfonts,amssymb}
\usepackage{graphicx,epsfig}
\usepackage{amsmath}
\usepackage{fancyhdr}
\usepackage{hyperref}
\usepackage{graphicx,epstopdf}
\hypersetup{
	colorlinks   = true, %Colours links instead of ugly boxes
	urlcolor     = blue, %Colour for external hyperlinks
	linkcolor    = blue, %Colour of internal links
	citecolor   = red %Colour of citations
}

\def\o{\omega}

\def\no{\nonumber}

\def\d{\delta}

\def\p{\partial}

\def\na{\nabla}
\def\T{\Theta}
\def\lie{\pounds_{\xi}}
\def\t{\tilde}
\def\tlie{\pounds_{\t{\xi}}}
\def\l{\lambda}
\begin{document}
\title{Noether and Abbott-Deser-Tekin conserved quantities in scalar-tensor theory of gravity both in Jordan and Einstein frames}
\author{Krishnakanta Bhattacharya\footnote{\color{blue} krishnakanta@iitg.ernet.in}}
\author{Ashmita Das\footnote {\color{blue} ashmita.phy@gmail.com}}
\author{Bibhas Ranjan Majhi\footnote {\color{blue} bibhas.majhi@iitg.ernet.in}}

\affiliation{Department of Physics, Indian Institute of Technology Guwahati, Guwahati 781039, Assam, India}

\date{\today}

%%%%%%%%%%%%%%%%%%%%%%%%%%%%%%%%%%%%%%%%%%%%%%%%
\begin{abstract}
We revisit the thermodynamic aspects of the scalar-tensor theory of gravity in the Jordan and in the Einstein frame. 
Examining the {\it missing links} of this theory carefully, we establish the thermodynamic descriptions from the conserved currents and potentials by following both the Noether and the Abbott-Deser-Tekin (ADT) formalism. With the help of conserved Noether current and potential, we define the thermodynamic quantities, which we show to be {\it conformally invariant}. Moreover, the defined quantities are shown to fit nicely in the laws of (the first and the second) black hole thermodynamics formulated by the Wald's method. We stretch the study of the conformal equivalence of the physical quantities in these two frames by following the ADT formalism. Our further study reveals that there is a connection between the ADT and the Noether conserved quantities, which signifies that the ADT approach provide the equivalent thermodynamic description in the two frames as obtained in Noether prescription. Our whole analysis is very general as the conserved  Noether and ADT currents and potentials are formulated {\it off-shell} and the analysis is exempted from any prior assumption or boundary condition.
\end{abstract}

%%%%%%%%%%%%%%%%%%%%%%%%%%%%%%%%%%%%%%%%

\maketitle

%%%%%%%%%%%%%%%%%%%%%%%%%%%%%%%%%%%%%%%%%%%%%%%%%%%%%%%%%%%%%%%%%%%%%%%%%%%%%%%
\section{Introduction}
To date the predictions of General Relativity (GR), as proposed by Einstein, are successfully verified by several experiments with enormous degree of precision. The discovery of gravitational wave in the year of 2016 \cite{Abbott:2016blz} has added another feather to the crown. Despite of the big success of this theory, several recent experiments \cite{Perlmutter:1998np, Riess:2001gk, Tonry:2003zg} reveal that the Einstein's General Relativity is not a complete theory.  
 In order to address the limitations of GR, several modified theories of gravity are proposed which has been the subjects of ardent research works for past few decades. Each of these modified GR theories have their own significance and provides distinct motivation to analyse them critically. Among various modified theories of gravity, the scalar-tensor theory is the most popular one for many reasons \cite{Callan:1985ia, EspositoFarese:2003ze, Elizalde:2004mq, Saridakis:2016ahq, Crisostomi:2016czh, Langlois:2017dyl}. This modified theory can be analysed with respect to the two frames, one is known as Jordan frame where the conventional Lagrangian of GR gets modified with the inclusion of the scalar field $\phi$. As a result, in the modified action, the Ricci scalar gets minimally coupled with the scalar field. This non-minimal coupling can be removed by the conformal transformation of the metric tensor along with the re-scaling  of the scalar field and, by the virtue of these transformations, one arrives to the another frame, known as the Einstein frame.

There are lot of arguments on the fact that which of these two frames can be considered as more physical one \cite{Faraoni:1999hp, ALAL} (also see the reviews \cite{Faraoni:1998qx, Faraoni:2010yi} to get more insights). There is another controversial aspect of this theory which states whether the conformal equivalence of the action in the two frames is merely a mathematical equivalence or this equivalence is also reflected in the dynamical \cite{Faraoni:2006fx, Saltas:2010ga, Capozziello:2010sc, Padilla:2012ze} and the underlying thermodynamic aspects as well \cite{Koga:1998un, Jacobson:1993pf, Kang:1996rj, Dehghani:2006xt, Sheykhi:2009vc} (also see the recent papers \cite{Steinwachs:2011zs, Kamenshchik:2014waa, Banerjee:2016lco, Pandey:2016unk, Ruf:2017xon}, which discusses on the equivalence of the two frames in the quantum level). There are a few unsolved issues such as, what are the explicit covariant expressions of the physical quantities (energy, entropy, temperature) and how they are connected in the two frames. Although, the expression of the entropy and the temperature is  widely accepted \cite{Jacobson:1993pf, Kang:1996rj} to some extent but, there is a controversy in the expression of the energy which can be used for the thermodynamic description in this theory. Most of the existing expressions of energy (or mass) as described  in literature are not conformally invariant \cite{Prain:2015tda, Faraoni:2015sja, Hammad:2016yjq}, whereas, the expressions of the entropy and the temperature are conformally invariant. This makes the physicists more puzzled as none of the existing energy, so far, can be used together with entropy and temperature for the thermodynamic description. In this regard, the question, which remains unsolved for a long time is what is the thermodynamic approach to define the mass (or energy). Thus, in order to resolve these issues, more investigation is required to provide satisfactory answers to these questions.

In our previous work \cite{Bhattacharya:2017pqc}, we have systematically developed the arguments to prove that all the thermodynamic quantities (For example: energy, entropy, temperature) must be equivalent in the two frames, without taking any prior assumption but, we could not formulate the exact covariant expression of the thermodynamic quantities using the method which we have followed in \cite{Bhattacharya:2017pqc}.
It is well-known that the conserved currents in a theory play an important role to understand the thermodynamic aspects of the corresponding theory. Therefore, in this work, we formulate the thermodynamic descriptions using the conserved quantities following the two different methods. One is the Noether prescription of defining the conserved currents and the potentials due to the diffeomorphism and, the other one is the ADT method of defining conserved currents in the presence of a Killing vector. In both the cases, the conserved quantities will be obtained \textit{off-shell}. Therefore, the expressions of the currents and potentials are also applicable for a generic null-surface as well, which are not derived using the equations of motion. Therefore, from the viewpoint of the emergent gravity paradigm, our analysis play a significant role.  Using the obtained conserved quantities, we shall define the covariant expressions of all the thermodynamic quantities, which will be shown to fit nicely in the (first and second) laws of black hole thermodynamics. Subsequently, we show that the thermodynamic quantities are exactly conformally invariant without using any prior assumptions or boundary conditions. Similar conclusion has also been made in the context of cosmological aspects of this theory where several physical quantities have been defined in an invariant way \cite{Jarv:2014hma,Jarv:2016sow,Karam:2017zno}.

Moreover, we obtain that the conserved currents in these two approaches (Noether and ADT) are connected to each other and we shall show the explicit connection between them. In addition, the ADT potentials in two frames are shown to be related with each other in the similar fashion like the Noether counter parts, establishing the equivalence of thermodynamic quantities, defined by the ADT potentials.  Thus, our work, provides a robust method to formulate the off-shell conserved quantities in two different approaches and resolves the ambiguities in the thermodynamic descriptions in the scalar-tensor theory which prevailed for the last few decades.

 The paper will be organized as follows. In the following section, we provide a brief description of the scalar-tensor theory from the action level. 
 In the next section, we formulate the off-shell method of defining the conserved Noether current and potential in the two frames due to the diffeomorphism. In the later section, we shall define the thermodynamic quantities in these two frames and prove the first law of BH thermodynamics in each frame following the Iyer-Wald formalism \cite{Wald:1993nt}. Then, we establish the conformal invariance of defined thermodynamic quantities in the two frames and subsequently the entropy increasing theorem (the second law) is also established. Then, in sec.(\ref{ADT}), we focus on another approach called off-shell ADT formalism in the two frames of the scalar-tensor theory and obtain the corresponding thermodynamic descriptions. 
 We shall conclude our analysis highlighting the major outcomes and its implications in sec.(\ref{CON}).
%%%%%%%%%%%%%%%%%%%%%%%%%%%%%%%%%%%%%%%%%%%%%%%%%%%%%%%%%%%%%%%%%%%%%%%%%
%%%%%%%%%%%%%%%%%%%%%%%%%%%%%%%%%%%%%%%%%%%%%%%%%%%%%%%%%%%%%%%%%%%%%%%%%%%%%%%%%
\section{Action in Jordan and Einstein frames: in a nutshell} \label{ACTION}
Let us start describing the scalar-tensor theory from the rudimentary level, i.e., from the action level. The action of the scalar-tensor theory in the Jordan frame is given as 
\begin{eqnarray}
&&\mathcal{A}=\int d^4x\sqrt{-g}L =\int d^4x\sqrt{-g} \frac{1}{16\pi}\Big(\phi R
\nonumber
\\
&&-\frac{\omega (\phi)}{\phi}g^{ab}\nabla_a\phi \nabla_b\phi -V(\phi)\Big)~.
\label{SJ}
\end{eqnarray}
In this frame the scalar field $\phi$, is non-minimally coupled with the Ricci-scalar and the Brans-Dicke parameter $\omega$, which is a generalized function of the scalar field $\phi$. Here, we consider only the gravitational action for our further analysis. We mention that in our analysis, the presence of  the external matter fields does not play any significant role in order to obtain the relations among the thermodynamic quantities in these two frames in the background of a scalar-tensor theory,

Earlier works suggest that the non-minimal coupling of the scalar field with the Ricci-scalar can be eradicated with the help of the conformal transformation which is given as,
\begin{align}
g_{ab}\rightarrow\tilde{g}_{ab}=\Omega^2g_{ab},\ \ \ \ \ \ \ \ \Omega=\sqrt{\phi} ~,
\label{GAB}
\end{align}
with the re-scaling of the scalar field 
\begin{align}
 \phi\rightarrow\tilde{\phi}\,\ {\textrm{with}}\,\ d\tilde{\phi}=\sqrt{\frac{2\omega+3}{16\pi}}\frac{d\phi}{\phi}~.
\label{PHI}
\end{align}
In several literature it is mentioned that by the virtue of the above two simultaneous transformations \eqref{GAB} and \eqref{PHI}, one arrives to the Einstein frame and the action in this frame can be written as,
\begin{eqnarray}
&&\tilde{\mathcal{A}}=\int d^4x\sqrt{-\tilde{g}}\tilde{L}
\nonumber
\\
&&=\int d^4x\sqrt{-\tilde{g}}[\frac{\tilde{R}}{16\pi}-\frac{1}{2}\tilde{g}^{ab}\tilde{\nabla}_a\tilde{\phi}\tilde{\nabla}_b\tilde{\phi}-U(\tilde{\phi})]~,
\label{SE}
\end{eqnarray}
where $U(\tilde{\phi}) = \frac{V(\phi)}{16\pi\phi^2}$.
But, in our earlier work \cite{Bhattacharya:2017pqc} we have investigated the exact relation of the Lagrangians in these two frames due to the conformal transformation and obtain,
\begin{align}
\sqrt{-\tilde{g}}\tilde{L}=\sqrt{-g} L-\frac{3}{16\pi}\sqrt{-g}\square\phi~.
\label{ACT}
\end{align}

Let us now discuss about the last term of the above equation, which we shall see later plays an important role in our main analysis. Note that this term is a total derivative term and contains second order time derivative of $\phi$. Therefore it creates issues in obtaining the equation of motion for the following reasons. Being a second order derivative term, one needs to fix simultaneously the field and its canonical momentum at the two end points in the least action formalism. Classically, if we fix arbitrarily both the parameters $\phi$ as well as the first order derivative of $\phi$ at the two boundary points, there may not exist a classical solution for the field $\phi$ consistent with the boundary conditions. Moreover, in general, we prefer that the action principle must obey the composition rule. This implies that, at the intermediate point, the first order derivative of $\phi$ has to be continuous but not necessarily to be a smooth function. It infers that the first order derivative of $\phi$ remains arbitrary (for instance, see a detailed discussion in page 241 of \cite{paddy}). Thus, in the classical regime, fixing both $\phi$ and its first order derivative simultaneously at the boundary is not admissible. Moreover, this prescription stems further problem in quantizing the theory as the simultaneous application of these two boundary conditions, indeed contradicts the uncertainty principle. It may be pointed out that this problem, however, is not new in the context of the general relativity (GR) as the same happens in case of the variations of Einstein-Hilbert action with respect to $g_{ab}$ to obtain Einstein's equations of motion. Two ways are usually being adopted to resolve such situation. One needs to either discard this total derivative term or add a judiciously chosen boundary term which cancels the unwanted terms, appearing in the variation of the original action. For example, in GR, the popular boundary term is the Gibbons-Hawking-York (GHY) boundary term. The similar can be done in the present situation as well. Addition of a precise GHY like boundary term in this case has also been adopted in this theory (for a discussion, see \cite{Bhattacharya:2017pqc}). But remember that such a choice is not unique as there may exist other term which also serves the same purpose (for GR case, see \cite{Charap:1982kn}). 
On the other hand, following the other argument, the boundary term (the last term of Eq. (\ref{ACT})) has usually been disregarded in the literature.

The prescription of neglecting the total derivative term at the level of finding equation of motion may be permissible; but such a term can be important in defining various physical quantities of the theory. Observe that the actions in two frames are same if one does not discard the last term of Eq. (\ref{ACT}). Thus, discarding of it influences an in-built inequivalence between these Lagrangians which can yield several deeper inequivalence even at the classical level (see \cite{Bhattacharya:2017pqc} for this particular issue). Therefore, we keep this term for our analysis. In this paper our motivation is to find the thermodynamic quantities in the two frames and establish a connection between them. It may be worthwhile to point out that a boundary term can contribute to thermodynamical description of gravity. 
Therefore, in terms of finding thermodynamic quantities and their relations in these two frames, surface terms in the action may play a very important role. Hence for a robust study of the thermodynamic description of the scalar-tensor theory, now onward we perform our further analysis by considering the most general form of Lagrangian of the following form:
\begin{align}
 L'=L-(3/16\pi)\square\phi~. \label{L'}
\end{align}
Our analysis is in the same line of Noether prescription by Wald in GR case \cite{Wald:1993nt}. In this discussion one considers the diffeomorphism invariant action as $\sqrt{-g}R$ without dropping or adding a boundary term in it to define thermodynamics. Adopting the same spirit, we also do not discard or include anything in the theory. We shall find that the thermodynamic quantities are well defined and equivalent in two frames which was usually sporadically stated in earlier analysis, thereby establishing the importance of retaining the boundary term. 

%%%%%%%%%%%%%%%%%%%%%%%%%%%%%%%%%%%%%%%%%%%%%%%%%%%%%%%%%%%%%%%%%%%%%%%%%%%%%%%%%%%%

\section{Conserved quantities in Noether prescription: OFF-Shell condition}
In this section, we shall obtain the Noether current and the Noether potential due to the diffeomorphism invariance of the Lagrangian.  
We start with the analysis in Jordan frame followed by Einstein frame in the later part and we emphasize on the point that our approach is very general in nature where we perform our mathematical analysis under the off-shell condition. 

\subsection{Jordan frame quantities}
The arbitrary variation of the action with the Lagrangian $L'$ in the Jordan frame yields
\begin{align}
\delta(\sqrt{-g}L')=\sqrt{-g}E_{ab}\delta{g^{ab}}+\sqrt{-g}E_{(\phi)}\delta{\phi}
\no
\\
+\sqrt{-g}\na_a\T'^a (q,\delta{q})~, \label{VAR1}
\end{align}
where $q\in \{g_{ab}, \phi\}$.
$E_{ab}=0$ and $E_{(\phi)}=0$, provide the equations of motion for the metric tensor $g_{ab}$ and the scalar field $\phi$ respectively and $\T'^a (q,\delta{q})$ is the boundary term. As we look for the off-shell Noether and ADT currents, we nowhere use the equations of motion while obtaining those quantities. The exact expressions of $E_{ab}$, $E_{(\phi)}$ and $\T'^a (q,\delta{q})$ are given by,
\begin{align}
 E_{ab}=\frac{1}{16\pi}[\phi G_{ab}+\frac{\omega}{2\phi}\nabla_i\phi\nabla^i\phi g_{ab}-\frac{\omega}{\phi}\nabla_a\phi\nabla_b\phi
\no 
\\
+\frac{V}{2}g_{ab}-\nabla_a\nabla_b\phi+\nabla_i\nabla^i\phi g_{ab}]~;
\no
\\
E_{(\phi)}=\frac{1}{16\pi}[R+\frac{1}{\phi}\frac{d\omega}{d\phi}\nabla_i\phi\nabla^i\phi +\frac{2\o}{\phi}\square\phi-\frac{dV}{d\phi}
\nonumber
\\
-\frac{\omega}{\phi^2}\nabla_a\phi \nabla^a\phi]~;
\no 
\\
\textrm{and} \ \ \ \ \ \ \ \ \ \ \ \ \ \ \ \ \ \ \ \ \ \ \ \ \ \ \ \ \ \ \ \ \ 
\no 
\\
\T'^a (q,\delta q)=\T^a (q,\delta q)-\frac{1}{16\pi}\Big\{\frac{3}{2}g^{ij}\d g_{ij}\p^a\phi
\no 
\\
-3g^{ia}\p^b\phi\d g_{ib}+3\p^a(\d\phi)\Big\}~,  \label{EXACTEXPJOR}
\end{align} 
where,
\begin{equation}
\T^a (q,\delta{q})=\frac{1}{16\pi}[-2g^{ab}\frac{\omega}{\phi}(\nabla_b\phi) \delta\phi +\phi \delta v^a-2(\nabla_b\phi)p^{iabd}\delta g_{id}]~.\label{theta1}
\end{equation}

Here, $G_{ab}=R_{ab}-\frac{1}{2}g_{ab}R$ is the Einstein tensor and
\begin{align}
\delta v^a=2p^{ibad}\nabla_b \delta g_{id}~;\ \ \ \ \ \ \ \ \ \ \ \ \ \
\no 
\\
p^{iabd}=\partial R/\partial R_{iabd}= (1/2)[g^{ib}g^{ad}-g^{id}g^{ab}]~.
\label{theta2}
\end{align}
 The terms within the curly brackets in the expression of $\T'^a(q, \d q)$ in \eqref{EXACTEXPJOR}, are originated from the variation of the $\square\phi$ term in $L'$. People usually do not take the contribution from these extra terms but, we show later, these extra terms in $\T'^a(q, \d q)$ play crucial role in the conformal invariance of the thermodynamic quantities in the two frames.

It is well known that in Einstein's gravity, one can define the conserved off-shell Noether current due to the fact that the covariant derivative of the Einstein tensor vanishes (see the project 8.1 of \cite{paddy}). In the scalar-tensor theory, we are able to find out a similar identity, which helps us to formulate the off-shell Noether current.

From the Eq. \eqref{EXACTEXPJOR}, we calculate $\na_bE^{ab}$, which is given as 
\begin{align}
\na_bE^{ab}=G^{ab}(\na_b\phi)-\frac{1}{2\phi}\frac{d \omega}{d\phi}(\na^a\phi)(\na^b\phi)(\na_b\phi)
\no 
\\
-\frac{\omega}{\phi}(\na^a\phi)\square\phi+\frac{1}{2}\frac{d V}{d\phi}(\na^a\phi)-\na_b\na^a\na^b\phi+\na^a\na_b\na^b\phi~. \label{NAEAB}
\end{align}
Using $\na_b\na^a\na^b\phi-\na^a\na_b\na^b\phi=R^{ab}\na_b\phi$ in the above equation \eqref{NAEAB} and using the expression of $E_{\phi}$ from \eqref{EXACTEXPJOR}, one finally obtains,
\begin{align}
\na_bE^{ab}=-\frac{1}{2}(\na^a\phi) E_{(\phi)}~. \label{RELEE}
\end{align}
The above relation shows the explicit connection between $E_{ab}$ and $E_{(\phi)}$ which is not intuitively expected by looking into the first two equations of eq.(\ref{EXACTEXPJOR}). This relation in turn helps us to find out the explicit value of the off-shell Noether current and potential in the Jordan frame.

 Due to the diffeomorphism $x^a\rightarrow x^a+\xi^a$, the off-shell change in the Lagrangian is given from \eqref{VAR1} as, 
\begin{align}
\lie (\sqrt{-g}L')=-2\sqrt{-g}E_{ab}\na^a\xi^b+\sqrt{-g}E_{(\phi)}\xi^a\na_a\phi
\no 
\\
+\sqrt{-g}\na_a\T'^a (q,\lie q)~, \label{PREJAB}
\end{align} 
where $\lie$ denotes the Lie variation.
The LHS of the Eq. \eqref{PREJAB} gives 
\begin{align}
\lie (\sqrt{-g}L')=L'\lie (\sqrt{-g})+\sqrt{-g}\lie L')\ \ \ \ \
\no 
\\
=\sqrt{-g}L'\na_a\xi^a+\sqrt{-g}\xi^a\na_a L'=\sqrt{-g}\na_a(L'\xi^a)~.\label{L_xi_1}
\end{align}
The contribution from the RHS of \eqref{PREJAB} can be written as,
\begin{align*}
-2\sqrt{-g}\na_a(E^{ab}\xi_b)+2\sqrt{-g}\xi_b\na_aE^{ab} 
\\
+\sqrt{-g}E_{(\phi)}\xi^a\na_a\phi+\sqrt{-g}\na_a\T'^a (q,\lie q)~.
\end{align*}
 Using the relation of \eqref{RELEE} in the above expression, the whole expression reduces to a total derivative term which is given as, $\sqrt{-g}\na_a[-2E^{ab}\xi_b+\T'^a (q,\lie q)]$. \\
 Thus, finally \eqref{PREJAB} gives,
\begin{align}
\na_a[L'\xi^a+2E^{ab}\xi_b-\T'^a (q,\lie q)]=0~.
\end{align}
Therefore, one can identify the term within the square bracket as a conserved quantity which is nothing but the Noether current due to the diffeomorphism. We denote it by $J'^a$, where,
\begin{align}
J'^a=L'\xi^a+2E^{ab}\xi_b-\T'^a (q,\lie q)~. \label{J'A}
\end{align}
The above expression of $J'^a$ can be further expressed as $J'^a=\na_b J'^{ab}$, where the anti-symmetric (off-shell) Noether potential is given as (see the appendix \ref{APPEN1} for detail discussion)
\begin{align}
J'^{ab}=\frac{1}{16\pi}[\nabla^a(\phi\xi^b)-\nabla^b(\phi\xi^a)]~. \label{JAB'}
\end{align}
These quantities play an important role in the study of black hole thermodynamics which we shall derive by following the Iyer-Wald formalism in the subsequent section.

For the sake of completeness, let us mention the expression of Noether potential in the Jordan frame when one consider the Lagrangian $L$ instead of $L'$ and it can be depicted as, 
\begin{align}
J^{ab}=\frac{1}{16\pi}\Big[\phi(\nabla^a\xi^b-\nabla^b\xi^a)+2\xi^a(\nabla^b\phi)-2\xi^b(\nabla^a\phi)\Big]~. \label{JAB}
\end{align}
The above relation in \eqref{JAB} is usually given as the Noether potential in the literature \cite{Bhattacharya:2017pqc, Koga:1998un}. However, the Noether potential in \eqref{JAB}, cannot be expressed as proportional to the Noether potential in the Einstein frame. On the other hand, we subsequently show that, the expression of the Noether potential, as given in \eqref{JAB'}, can be written as proportional to the Noether potential in the Einstein frame, which imply conformal invariance of the thermodynamic quantities in these two frames.  
Let us now follow the Noether prescription in the Einstein frame.

%%%%%%%%%%%%%%%%%%%%%%%%%%%%%%%%%%%%%%%%%%%%%%%%%%%%%%%%%%%%%%%%%%%%%%%%%%%%%%%
\subsection{Einstein frame quantities}
Applying the variational principle on the action mentioned in Eq. \eqref{SE}, we obtain 
\begin{align}
\d(\sqrt{-\tilde{g}}\t{L})=\sqrt{-\t{g}}\t{E}_{ab}\delta{\t{g}^{ab}}+\sqrt{-\t{g}}\t{E}_{(\t{\phi})}\delta{\t{\phi}}
\no
\\
+\sqrt{-\t{g}}\t{\na}_a\t{\T}^a (\t{q},\delta{\t{q}})~, \label{VAR2}
\end{align}
where $\t{q}\in \{\t{g}_{ab}, \t{\phi}\}$ and
\begin{align}
 \t{E}_{ab}=\frac{\tilde{G}_{ab}}{16\pi}-\frac{1}{2}\tilde{\nabla}_a\tilde{\phi}\tilde{\nabla}_b\tilde{\phi}+\frac{1}{4}\tilde{g}_{ab}\tilde{\nabla}^i\tilde{\phi}\tilde{\nabla}_i\tilde{\phi}
\nonumber
\\
+\frac{1}{2}\tilde{g}_{ab}U(\tilde{\phi})~;
\no 
\\
\t{E}_{(\t{\phi})}=\t{\na}_a\t{\nabla}^a\tilde{\phi}-\frac{dU}{d\tilde{\phi}}~;\ \ \ \ \ \ \ \ \ \ \ \
\no 
\\
\textrm{and} \ \ \ \ \ \ \ \ \ \ \ \ \ \ \ \ \ \ \ \ \ \ \ \ \
\no 
\\
\t{\T}^a (\t{q},\delta{\t{q}})=\frac{\delta\tilde{v}^a}{16\pi}-(\tilde{\nabla}^a\tilde{\phi})\delta\tilde{\phi}~. \ \ \ \ \ \  \label{EXACTEXEIN}
\end{align}
As in the earlier case, here $\t{G}_{ab}$ is the Einstein tensor in this frame and
\begin{align}
\delta \t{v}^a=2\t{p}^{ibad}\t{\nabla}_b \delta \t{g}_{id};\ \ \ \ \ \
\no 
\\
\t{p}^{iabd}= (1/2)[\t{g}^{ib}\t{g}^{ad}-\t{g}^{id}\t{g}^{ab}]~.
\end{align}
Proceeding similarly as in the Jordan frame analysis, here also we work under the off-shell condition and in order to define off-shell conserved quantities one needs a Bianchi-type identity in the Einstein frame. From \eqref{EXACTEXEIN}, it is straightforward to show that a similar expression as in the Jordan frame [{\it i.e}\eqref{RELEE}] can be obtained as follows,
\begin{align}
\t{\na}_b\t{E}^{ab}=-\frac{1}{2}(\t{\na}^a\t{\phi})\Big[\t{\square}\t{\phi}-\frac{dU}{d\t{\phi}}\Big]=-\frac{1}{2}(\t{\na}^a\t{\phi})\t{E}_{\phi}~. \label{RELEEE}
\end{align}
Like the Jordan frame analysis, here we use the above equation
to derive the off-shell Noether current and Noether potential in the Einstein frame.
We refer our reader to the earlier mathematical analysis as done in the Jordan frame and Appendix(\ref{APPEN3}) in order to get a detail calculations of the Noether current and Noether potential in Einstein frame. Hence, here we summarise our result for Noether current and potential as,
\begin{align}
\t{J}^a=\t{L}\t{\xi}^a+2\t{E}^{ab}\t{\xi}_b-\t{\T}^a (\t{q},\lie \t{q})~. \label{JEINT}
\end{align}
and,
\begin{align}
\t{J}^{ab}=\frac{1}{16\pi}[\tilde{\nabla}^a\tilde{\xi}^b-\tilde{\nabla}^b\tilde{\xi}^a]~. \label{JABEIN}
\end{align}
Thus, we obtain the conserved off-shell Noether current and the Noether potential in the two frames. 

Now, we adopt the Wald's formalism to established the first law of the black hole thermodynamics in the two frames by using the above derived quantities, in the following section. We shall define all the thermodynamic quantities and show the explicit conformal invariance of these quantities in the two frames in the background of the scalar-tensor theory.
%%%%%%%%%%%%%%%%%%%%%%%%%%%%%%%%%%%%%%%%%%%%%%%%%%%%%%%%%%%%%%%%%%%%%%%%%%%%%%%%%
\section{Thermodynamic quantities by Iyer-Wald formalism and their conformal invariance} \label{WALD}
\subsection{Jordan frame}
The expression of the Noether current and the Noether potential in the Jordan frame are given in \eqref{J'A} and \eqref{JAB'}. According to the Wald's formalism we shall use the on-shell condition which is given by, $E^{ab}=0$. Let us now take the variation of the metric tensor and the scalar field which leaves the diffeomorphism vector $\xi^a$ invariant (remember, $\d\xi_a\neq 0$ in general as $\d g_{ab}\neq 0$) and therefore the change in the conserved on-shell Noether current with respect to the variation of the fields becomes,
\begin{align}
\delta(\sqrt{-g}J'^a)=\d(\sqrt{-g}L')\xi^a-\d [\sqrt{-g}\T'^a(q, \lie q)]~. \label{DELJ}
\end{align}
Using Eq. \eqref{VAR1}, we get the variation of the Noether current in terms of the boundary term $\T'^a$, which is given as
\begin{align}
\delta(\sqrt{-g}J'^a)=\sqrt{-g}[\na_i\T'^i(q, \d q)]\xi^a-\d [\sqrt{-g}\T'^a(q, \lie q)]~. \label{VAR}
\end{align}
We shall see that this variation of the Noether current can be written in terms of the symplectic Hamiltonian density by using an identity which one can straightforwardly obtain: 
\begin{align}
\pounds_{\xi}[\sqrt{-g} \T'^a(q, \d q)]=\sqrt{-g}\xi^a\nabla_i[\T'^i(q, \d q)]
\no 
\\
-2\sqrt{-g}\nabla_b[\xi^{[a}\T'^{b]}(q, \d q)]~,
\end{align}
where $A^{[a}B^{b]}=(1/2)(A^aB^b-A^bB^a)$. Using the above identity in \eqref{VAR}, we obtain
\begin{align}
\delta(\sqrt{-g}J'^a)=\pounds_{\xi}[\sqrt{-g} \T'^a(q, \d q)]-\d [\sqrt{-g}\T'^a(q, \lie q)]
\no 
\\
+2\sqrt{-g}\nabla_b[\xi^{[a}\T'^{b]}(q, \d q)]~. \label{VARJ}
\end{align}
Now define:
\begin{align}
\o^a=-\pounds_{\xi}[\sqrt{-g} \T'^a(q, \d q)]+\d [\sqrt{-g}\T'^a(q, \lie q)]~. \label{O1}
\end{align} 
 The significance of $\o^a$ will be explained in a few steps later. With this definition of $\o^a$, one can obtain from \eqref{VARJ},
\begin{align}
\omega^a=-\delta(\sqrt{-g}J'^a)+2\sqrt{-g}\nabla_b[\xi^{[a}\T'^{b]}(q, \d q)].  \label{O2}
\end{align}
Let us now discuss the significance of $\o^a$. For a classical system we write, $\delta L(x_i, \dot{x_i})=[(\frac{\partial L}{\partial x_i})-d_t(\frac{\partial L}{\partial \dot{x_i}})]\delta x_i +d_t[p^i\delta x_i]$ where, $x_i$ is the generalized coordinate and $p^i=\frac{\partial L}{\partial \dot{x_i}}$ is the generalized momentum. The equation of motion vanishes on-shell and, the variation of the Hamiltonian due to the arbitrary variation of the the coordinates $x_i$ is given as 
\begin{align}
\delta H(x_i,p^i)=\delta[p^i(d_tx_i)]-d_t[p^i(\delta x_i)]~. \label{DELH}
\end{align}
By comparing \eqref{O1} and \eqref{DELH}, one can identify that $\o^a$ as the variation of the symplectic Hamiltonian density where the boundary terms in both the equations are equivalent to each other as,  $\sqrt{-g} \T'^a(q, \d q)\equiv p^i(\delta x_i)$ and $\sqrt{-g}\T'^a(q, \lie q)\equiv p^i(d_tx_i)$.

 Thus, with this above identification, the total variation of the Hamiltonian can be written as (using the Eq. \eqref{O2}),
\begin{eqnarray}
&&\delta H[\xi]= \int_c d\Sigma_a \frac{\omega^a}{\sqrt{-g}}
\nonumber
\\
&&=-\d\int_cd\Sigma_a\nabla_b(J'^{ab})+2\int_c d\Sigma_a\nabla_b[\xi^{[a}\T'^{b]}(q, \d q)]~, 
\no 
\\
\ \ \label{DELH1}
\end{eqnarray} 
where, the integration is done on Cauchy hypersurface which we symbolize as $c$. $d\Sigma_a=n_a\sqrt{h}d^3x$ is the elemental surface area of the three-dimensional Cauchy hypersurface, with $n_a$ being the normal and $h$ being the determinant of the induced metric of the surface. Applying the Stoke's law in the above equation we can reduce the 3-surface integral of above \eqref{DELH1} to a 2-surface integral. We consider $\xi^a$ is a Killing vector and the outer surface lies at assymptotic infinity ({\it i.e} $\p c_{\infty}$). The inner surface of $c$ is taken as a bifurcation surface {\it i.e} $\mathcal{H}$ which also can be depicted as the horizon of the black hole. This implies $\xi^a=0$ at  $\mathcal{H}$. 
Thus, from \eqref{DELH1} we obtain,
\begin{align}
\delta H[\xi]=-\frac{1}{2}\delta\int_{\mathcal{H}} d\Sigma_{ab}J'^{ab}+\frac{1}{2}\delta\int_{\partial c_{\infty}} d\Sigma_{ab}J'^{ab}
\nonumber
\\
-\int_{\partial c_{\infty}} d\Sigma_{ab} \xi^{[a}\T'^{b]}(q, \d q)~. \label{DELH2}
\end{align}
As $\xi^a=0$, no contribution comes from the term $\xi^{[a}\T'^{b]}(q, \d q)$ on $\mathcal{H}$. Moreover, as $\xi^a$ is a Killing vector, $\delta H[\xi]=0$. By following the Wald's prescription \cite{Wald:1993nt}, the first term on the RHS of \eqref{DELH2},  yields $-\frac{\kappa}{2\pi}\d S$ with $\kappa$ being the surface gravity and, the other terms result in $\d M-\Omega_H\d J$ (for a more rigorous discussions see \cite{Wald:1993nt}). Here, we define the entropy ($S$), the mass of the black hole ($M$), and the angular momentum ($J$) as,
\begin{eqnarray}
&& \d S=\frac{\pi}{\kappa}\delta\int_{\mathcal{H}} d\Sigma_{ab}J'^{ab}~;
\no 
\\
&& \d M=\frac{1}{2}\int_{\partial c_{\infty}} [\d(d\Sigma_{ab}J'^{ab})-2d\Sigma_{ab}\xi^{[a}\T'^{b]}(q, \d q)]\Big|_{\xi=\xi_{(t)}}~;
\no 
\\
&& \d J=-\frac{1}{2}\int_{\partial c_{\infty}} [\d(d\Sigma_{ab}J'^{ab})-2d\Sigma_{ab}\xi^{[a}\T'^{b]}(q, \d q)]\Big|_{\xi=\xi_{(\phi)}}~.
\no 
\\
 \label{SMJ}
\end{eqnarray}
So, finally from \eqref{DELH2}, we obtain
\begin{align}
\d M=T\d S+\Omega_H \d J~, \label{1STLAW}
\end{align}
where we use temperature $T=\kappa/(2\pi)$ in the above equation.
We comment that eq. \eqref{1STLAW} is the desired form of first law of the black hole thermodynamics in the Jordan frame with the Lagrangian $L'$. Instead of $L'$, if one consider the Lagrangian as $L$ in the Jordan frame, the expression of the entropy, mass and the angular momentum of the black hole can be obtained by replacing $J'^{ab}$ with $J^{ab}$ and $\T'^{b}(q, \d q)$ with $\T^{b}(q, \d q)$ in \eqref{SMJ}.
Let us now approach toward the Einstein frame and find out the thermodynamic quantities in that frame. 

\subsection{Einstein frame}
Proceeding similarly as the analysis of the first law of thermodynamics in the Jordan frame in previous subsection, it takes hardly any computation to affirm that in the Einstein frame we get the first law of the black hole mechanics as $\d \t{M}=\t{T}\d \t{S}+\t{\Omega}_H \d \t{J}$ and, the corresponding thermodynamic quantities are defined as 
\begin{eqnarray}
&& \d \t{S}=\frac{\pi}{\t{\kappa}}\delta\int_{\mathcal{H}} d\t{\Sigma}_{ab}\t{J}^{ab}~;
\no 
\\
&& \d \t{M}=\frac{1}{2}\int_{\partial c_{\infty}}[\d( d\t{\Sigma}_{ab}\t{J}^{ab})-2d\t\Sigma_{ab}\t{\xi}^{[a}\t{\T}^{b]}(\t{q}, \d \t{q})]\Big|_{\t{\xi}=\t{\xi}_{(t)}}~;
\no 
\\
&& \d \t{J}=-\frac{1}{2}\int_{\partial c_{\infty}} [\d(d\t{\Sigma}_{ab}\t{J}^{ab})-2d\t\Sigma_{ab}\t{\xi}^{[a}\t{\T}^{b]}(\t{q}, \d \t{q})]\Big|_{\t{\xi}=\t{\xi}_{(\phi)}}~.
\no 
\\
 \label{SMJTIL}
\end{eqnarray}
Let us now compare the thermodynamic quantities obtained in the two frames.
%%%%%%%%%%%%%%%%%%%%%%%%%%%%%%%%%%%%%%%%%%%%%%%%%%%%%%%%%%%%%%%%%%%%%%%%%%%%%
\subsection{Comparison of the thermodynamic quantities:}
 We consider the Killing vector in the Einstein frame ($\t{\xi^a}$) is same as in the Jordan frame i.e., $\t{\xi^a}=\xi^a$. The justification of taking the Killing vectors $\t{\xi^a}=\xi^a$ can be found in \cite{Bhattacharya:2017pqc}. The idea is the following. If $\xi^a$ is a Killing vector in Jordan frame, then it must be a conformal Killing vector in Einstein frame (see \cite{Jacobson:1993pf} for a discussion on this under conformal transformation). Remember that here we are discussing the whole thermodynamics in presence of Killing vector in both frames. Therefore $\tilde{\xi}^a=\xi^a$ to be Killing one, we need to impose the condition that the conformal factor must be Lie transported along $\xi^a$; i.e. $\pounds_\xi \Omega^2=0$.
 Earlier the authors in \cite{Koga:1998un} have addressed this issue by assuming the above condition and shown that the thermodynamic quantities are equivalent in these two frames under the condition of spacetime to be asymptotically flat. 

As $\t{\xi^a}=\xi^a$, we obtain $\t{\xi_a}=\phi\xi_a$ and, the relation between the complimentary null vectors in the two frames are given as $l_a=\t{l}_a$. Thus, 
\begin{align}
d\t{\Sigma}_{ab}=\sqrt{\t{\sigma}}(\t{\xi}_a\t{l}_b-\t{\xi}_b\t{l}_a)d^2x=\phi^2d\Sigma_{ab}~,
\end{align}
where $\sigma$ and $\t{\sigma}=\phi^2\sigma$ are the determinant of the induced metric of the two-surface in the Jordan and Einstein frames respectively. Therefore using the above relation, it can be easily seen that, (see the appendix \ref{APPEN4} for detail discussion)
\begin{align}
\t{J}^{ab}=\frac{J'^{ab}}{\phi^2}~. \label{JABJAB'}
\end{align}
In the appendix \ref{APPEN4}, we also show that,
\begin{align}
\t{\Theta^a}=\frac{\Theta'^a}{\phi^2}~. \label{THTH'}
\end{align}
Using the above relations, it can be seen that $\t{S}=S$, $\t{M}=M$ and $\t{J}=J$ in these two frames.  We comment that the equivalence of the angular velocity and the surface gravity (or the temperature) in these two frames can be shown by following the procedure as described in \cite{Koga:1998un}. 

We want to emphasize on the fact that $J^{ab}$ and $\Theta^a$ in the Jordan frame (when one takes the Lagrangian as $L$ instead of $L'$), cannot be written as proportional to the corresponding quantities in the Einstein frame. Therefore, one cannot establish the exact equivalence of the thermodynamic quantities between the Jordan and the Einstein frame, by considering the Lagrangian $L$ in the Jordan frame. Whereas, in our case, we show the conserved Noether potentials of the two frames are proportional to each other with the proportionality factor as $\phi^2$. This implies, in our case, the conserved Noether charge is the same in two frames.
We want to further emphasize that in the work of Koga and Maeda \cite{Koga:1998un}, assuming the spacetime to be asymptotically flat, the equivalence of the thermodynamic quantities in the two frames have been established by following the Wald's formalism. On the contrary in our work, by considering a more generalised Lagrangian $L'$, we establish the exact equivalence of thermodynamic parameters without making any assumption or imposing boundary conditions. 
 Therefore, in this regard our analysis is more general and implying a crucial fact that in order to explore the thermodynamic equivalence in the two frames, one needs to consider the Lagrangian as $L'$ in the Jordan frame instead of $L$.  
%%%%%%%%%%%%%%%%%%%%%%%%%%%%%%%%%%%%%%%%%%%%%%%%%%%%%%%%%%%%%%%%%%%%%%%%%%%%%
\subsection{Connection of the  derived mass with the Brown-York mass term}
Above, we have defined the masses in the two frames which are conformally invariant and are compatible with the first law. In literature, there are several prescription of defining the mass but, most of them are not conformally invariant. The only candidate, which is conformally invariant in the literature, is the Brown-York (BY) mass \cite{Brown:1992br} (also see \cite{Bose:1998yp} which discusses that the BY mass is conformally invariant but, the BY energy is not). Therefore, we investigate whether the derived expressions of mass in \eqref{SMJ} and \eqref{SMJTIL} are the same as the BY mass. Here, we do the analysis in the Einstein frame for simplicity. From the transformation relations of the quantities, the same conclusion can be drawn in the Jordan frame as well.

We consider 2-dimensional null-hypersurface characterised by the induced metric $\t\sigma_{ab}=\t g_{ab}-\t n_a\t n_b+\t u_a \t u_b$, where $\t u^a$ and $\t n^a$ are the timelike and spacelike normals respectively. From the above expression of $\d\t M$ in \eqref{EXACTEXEIN} we obtain,
\begin{align}
\d\t M=\d \t M_{BY}-\frac{1}{8\pi}\int d^2\t x\d(\sqrt{\t h}\t K^{(3)})
\no 
\\
+\int d^2\t x[\sqrt{\t h}\t n_a\t\T^a(\t q,\d\t q)]~, \label{BYORK}
\end{align}
where, $\t M_{BY}=\frac{1}{8\pi}\int d^2\t x\t N\sqrt{\t\sigma}\t k^{(2)}$ is the expression for BY mass with $\t N$ being the lapse function and $\t k^{(2)}$ being the trace of the extrinsic curvature tensor of the null surface and $\t K^{(3)}=\t\na_a\t n^a$ is the trace of the extrinsic curvature tensor of the induced 3-surface characterised by the induced metric $\t h_{ab}=\t g_{ab}-\t n_a\t n_b$. The above relation \eqref{BYORK} shows the explicit connection of our derived mass with the BY mass. The above relation can be further modified using eq. (12.104) of \cite{paddy}, which is given as
\begin{align}
\d\t M=\d \t M_{BY}-\frac{1}{16\pi}\int d^2\t x\sqrt{\t h}\Big[\Big(\t K_{ab}^{(3)}-\t K^{(3)}\t h_{ab}\Big)\d\t h^{ab}
\no 
\\
-\t D_i\t U^i+\t n^a(\t\na_a\t\phi)\d\t\phi\Big]~.
\end{align}
Here, $\t D_i$ denotes the covariant derivative operator in the three-space $\t h_{ab}$ and $\t U^i=2\t n_j\t h^i_k\d \t g^{jk}-\t n^i\t h_{jk}\t g^{jk}$. This shows that our mass is connected with the BY mass with some additive terms. 
%%%%%%%%%%%%%%%%%%%%%%%%%%%%%%%%%%%%%%%%%%%%%%%%%%%%%%%%%%%%%%%%%%%%%%%%%%%%%%
\subsection{Entropy increase theorem and the modified null energy condition in Jordan frame}
We analyse the entropy increase theorem in the background of this framework in order to get a complete picture of thermodynamic description of the scalar-tensor theory. Usually in GR the entropy increase theorem is established by assuming the null energy condition. But, we do not know what would be the null energy condition in the Jordan frame. Hence, one has to search for a similar energy condition which is different from the usual null energy condition. Here we show an interesting fact that the obtained similar energy condition in the Jordan frame, is proportional to the null energy condition as defined in the Einstein frame.

In this context a similar work has been done in \cite{Chatterjee:2012zh}, where the authors have interpreted the term at the right hand side (RHS) of the eq.(5) of \cite{Chatterjee:2012zh} as the stress-energy tensor of the scalar field $\phi$. But, we have not adopted that approach in our analysis. In our work, using $E_{ab}=0$ from \eqref{EXACTEXPJOR}, we obtain,
\begin{align}
G_{ab}=-\frac{\omega}{2\phi^2}\nabla_i\phi\nabla^i\phi g_{ab}+\frac{\omega}{\phi^2}\nabla_a\phi\nabla_b\phi-\frac{V}{2\phi}g_{ab}
\no    
\\
+\frac{1}{\phi}\nabla_a\nabla_b\phi-\frac{1}{\phi}\nabla_i\nabla^i\phi g_{ab}~.
\label{G_AB1}
\end{align}
From the above equation we cannot identify the RHS as the energy-momentum (EM) tensor of the scalar field $\phi$ in the Jordan frame as this is not compatible with the usual definition of the EM tensor (given as $T_{ab}=\frac{2}{\sqrt{-g}}\frac{\d L_{matter}}{\d g^{ab}}$). Thus, in this section, we try to provide a justifiable way to obtain the increase in the entropy by using the modified energy condition. 

From \eqref{G_AB1}, we calculate $R_{ab}l^al^b$ (with $l^a$ being a null vector), which is given as,
\begin{align}
R_{ab}l^al^b=\frac{\o}{\phi^2}(l^a\na_a\phi)^2+\frac{1}{\phi}l^al^b\na_a\na_b\phi~.
\end{align}
The first term is a positive definite for $\o>0$. Thus we write
\begin{align}
R_{ab}l^al^b-\frac{1}{\phi}l^al^b\na_a\na_b\phi\geq 0~. \label{ENCON}
\end{align}
The expression of the entropy in the Jordan frame is given in \eqref{SMJ} and using this equation our explicit calculation shows that the entropy can be written as $S=A/4$, where
\begin{align}
A=\int_{\mathcal{H}} \sqrt{\sigma}\phi d^2x~.
\end{align} 
The above expression of the entropy matches to the Kang's prescription in \cite{Kang:1996rj}.
Let us now find out the change in entropy along a null geodesic congruence. Hence, we calculate
\begin{align}
\frac{dA}{d\l}=\int_{\mathcal{H}} \sqrt{\sigma}\phi d^2x\theta' ~.
\end{align}
Here, $\l$ parametrizes the null-congruence and $\theta'=\theta^{(l)}+\frac{1}{\phi}\frac{d\phi}{d\l}$~, where $\theta^{(l)}=\frac{1}{\sqrt{\sigma}}\frac{d\sqrt{\sigma}}{d\lambda}$ is the expansion parameter along the null vector $l^a$. We intend to establish in the following analysis that $\frac{dS}{d\l}\geq$ always, by showing $\T'\geq0$. 
\begin{align}
\frac{d\theta'}{d\l}=\frac{d\theta^{(l)}}{d\l}-\frac{1}{\phi^2}(l^a\na_a\phi)^2+\frac{1}{\phi}l^al^b(\na_a\na_b\phi)~,
\no 
\\
=-\frac{1}{2}\theta^2-\sigma^2-R_{ab}l^al^b-\frac{1}{\phi^2}(l^a\na_a\phi)^2+\frac{1}{\phi}l^al^b(\na_a\na_b\phi)~.
\end{align}
The last expression is obtained using null Raychaudhuri equation where the null vector $l^a$ is an orthogonal-hypersurface. Using \eqref{ENCON} we obtain $\frac{d\theta'}{d\l}\leq0~.$ Therefore, the prohibition of caustics demands that $\theta'\geq0$. Thus the entropy increase theorem is established in this frame.

We now discuss that what is the significance of the condition in \eqref{ENCON}. Although \eqref{ENCON} is an identity in the Jordan frame, here we urge to prove that it corresponds to the null energy condition in the Einstein frame. 

In the Einstein frame,
\begin{align}
\frac{\t G_{ab}}{16\pi}=\frac{1}{2}\tilde{\nabla}_a\tilde{\phi}\tilde{\nabla}_b\tilde{\phi}-\frac{1}{4}\tilde{g}_{ab}\tilde{\nabla}^i\tilde{\phi}\tilde{\nabla}_i\tilde{\phi}-\frac{1}{2}\tilde{g}_{ab}U(\tilde{\phi})~.
\end{align}
 The right hand side of the above equation can be identified  as the stress-energy tensor ($\frac{\t T_{ab}^{(\t\phi)}}{2}$) of the scalar field $\t\phi$. Thus we obtain
\begin{align}
\t T_{ab}^{(\t\phi)}l^al^b=(\t l^a\t\na_a\t\phi)^2\geq0~. \label{ENCONEIN}
\end{align}
The above equation \eqref{ENCONEIN} is the null energy condition in the Einstein frame.  
Due to the conformal transformation we obtain,
\begin{align}
\t T_{ab}^{(\t\phi)}l^al^b=\frac{1}{\phi^2\o}\Big(\frac{2\o+3}{16\pi}\Big)\Big[R_{ab}l^al^b-\frac{1}{\phi}l^al^b\na_a\na_b\phi\Big]~.
\end{align}
Thus, we can conclude that the energy condition in the Jordan frame \eqref{ENCON} corresponds to the null energy condition in the Einstein frame.
%%%%%%%%%%%%%%%%%%%%%%%%%%%%%%%%%%%%%%%%%%%%%%%%%%%%%%%%%%%%%%%%%%%%%%%%%%%%%
\section{Off-shell ADT potential} \label{ADT}
The identification of the conserved charges in GR has always been an important task for decades. There are several methods of defining the conserved charges, each with some advantages and disadvantages in its way. The ADM formalism \cite{Arnowitt:1962hi} of computing the total conserved charge due to the Killing vectors has enjoyed the central attention, which holds good for the asymptotically flat spacetime. However, for the asymptotically non-flat or AdS spacetime, this approach fails.

For the asymptotically AdS solutions, a covariant method was developed by Abbott and Deser \cite{Abbott:1981ff} to compute the conserved Killing charges asymptotically. This method was later extended by Deser and Tekin for the higher order gravity theories \cite{Deser:2002rt, Deser:2002jk, Deser:2003vh} which popularly known as the Abbott-Deser-Tekin (ADT) formalism. Here, we extend the ADT formalism in the scalar-tensor theory which is absent in literature. Moreover, we show the explicit connection between the off-shell Noether potential and the ADT potential and address the issue of invariance of the ADT potentials in these two frames. 

For a Killing vector $\xi^a$, one can write,
\begin{align}
J^i_{ADT}|_{on-shell}=\d E^{ij}\xi_j~, \label{JADT}
\end{align}
which indeed is a conserved quantity under the on-shell condition. Here, $\d E^{ij}$ is the linearized tensor (first order change in the Equation of motion of the metric tensor due to $g^{ab}\rightarrow g^{ab}+\d g^{ab}$). The conservation of the $J^i_{ADT}$ follows from the fact that $\na_b\d E^{ab}=0$ on-shell (using eq (\ref{RELEE}))  and the property of the Killing vector (i.e. $\d E^{ab}\na_a\xi_b=0$). This conserved current we call as the ADT current. In the similar manner, the conserved on-shell ADT current in the Einstein frame can be written as, $\t J^i_{ADT}|_{on-shell}=\d \t E^{ij}\t\xi_j$. At this stage, we urge to derive the off-shell ADT currents in order to make a more general and robust analysis. Hence, we define the off-shell ADT currents in each frame and follow the similar method as done in Einstein's gravity case \cite{Bouchareb:2007yx}.

\subsection{Jordan frame}
We obtain that off-shell $\d E^{ij}\xi_j$ can be written as an anti-symmetric total derivative term added with some extra terms, where each of the extra terms is proportional to the $E_{ab}$ i.e.,  
\begin{align}
\d E^{ij}\xi_j=\na_jJ^{ij}_{ADT}-E^{ik}h_{kj}\xi^j+\frac{1}{2}\xi^iE^{jk}h_{jk}-\frac{1}{2}\xi^jE^i_jh~, \label{DeltaE}
\end{align}
where
\begin{eqnarray}
&&J^{ij}_{ADT}=\frac{1}{32\pi}\Big[\phi\Big(\xi^j\na_k h^{ki}-\xi^i\na_k h^{kj}+\xi_k\na^i h^{kj}
\no 
\\
&&-\xi_k\na^j h^{ki}+\xi^i(\na^jh)-\xi^j(\na^ih)+h^{jk}\na_k\xi^i-h^{ik}\na_k\xi^j
\no 
\\
&&+h\na^{[i}\xi^{j]}\Big)+(\na_k\phi)\Big(\xi^jh^{ik}-\xi^ih^{jk}\Big)\Big]~. \label{JIJADT}
\end{eqnarray}
and $h_{ab}=\d g_{ab}$ or equivalently $h^{ab}=-\d g^{ab}$. Identifying $J^i_{ADT}=\na_jJ^{ij}_{ADT}$ in eq.(\ref{DeltaE}), we obtain
%Shifting the extra terms on the left hand side, we can define the off-shell ADT current as 
\begin{align}
J^i_{ADT}|_{off-shell}=\d E^{ij}\xi_j+E^{ik}h_{kj}\xi^j-\frac{1}{2}\xi^iE^{jk}h_{jk}
\no 
\\
+\frac{1}{2}\xi^jE^i_jh~. \label{JADTOFF}
\end{align}
We refer our readers to the appendix \ref{APPEN5} for the detail derivation.

As $J^{ij}_{ADT}$ is an anti-symmetric tensor, $\na_iJ^{i}_{ADT}=0$ even in the off-shell which imply that the off-shell ADT current is also a conserved quantity. 

We now find out the conserved ADT current and potential in the Einstein frame in the following section.
%%%%%%%%%%%%%%%%%%%%%%%%%%%%%%%%%%%%%%%%%%%%%%%%%%%%%%%%%%%%%%%%%%%%%%%%%%%%
\subsection{Einstein frame}
As similar to the Jordan frame, off-shell $\d \t E^{ij}\t\xi_j$ can be written as, 
\begin{align}
\d \t E^{ij}\t\xi_j=\t\na_j\t J^{ij}_{ADT}-\t E^{ik}\t h_{kj}\t\xi^j+\frac{1}{2}\t \xi^i\t E^{jk}\t h_{jk}-\frac{1}{2}\t \xi^j\t E^i_j\t h~, \label{DeltaE2}
\end{align}
where
\begin{align}
\t J^{ij}_{ADT}=\frac{1}{32\pi}\Big[\t \xi^j\t\na_k \t h^{ki}-\t \xi^i\t \na_k \t h^{kj}+\t\xi_k\t\na^i \t h^{kj}-\t \xi_k\t\na^j \t h^{ki}
\no 
\\
+\t\xi^i(\t\na^j\t h)-\t\xi^j(\t\na^i\t h)+\t h^{jk}\t\na_k\t\xi^i-\t h^{ik}\t\na_k\t\xi^j+\t h\t\na^{[i}\t\xi^{j]}\Big]~, \label{JIJADTTIL}
\end{align}
and  $\t h_{ab}=\d \t g_{ab}$, $\t h^{ab}=-\d \t g^{ab}$.
Therefore following the same analogy as in Jordan frame, one can define the off-shell ADT current in the Einstein frame as, 
\begin{align}
\t J^i_{ADT}|_{off-shell}=\d \t E^{ij}\t\xi_j+\t E^{ik}\t h_{kj}\t\xi^j-\frac{1}{2}\t \xi^i\t E^{jk}\t h_{jk}
\no 
\\
+\frac{1}{2}\t \xi^j\t E^i_j\t h~. \label{JADTOFFTIL}
\end{align}
We refer the appendix \ref{APPEN6} for the detail derivation of the above equation \eqref{JIJADTTIL}. 
Following the earlier arguments, we comment that also in the Einstein frame the off-shell ADT current {\it i.e} $\t J^i_{ADT}$ is a conserved quantity.
%%%%%%%%%%%%%%%%%%%%%%%%%%%%%%%%%%%%%%%%%%%%%%%%%%%%%%%%%%%%%%%%%%%%%%%%%%%
\section{Connection between conserved off-shell ADT and Noether potentials} 
Here, we urge to study the connection between the off-shell ADT potential and the Noether potential. For the Einstein's gravity, this connection has been studied in literature \cite{Kim:2013zha}. The equation \eqref{JADTOFF} can be written as, (we drop the the subscript ``off-shell" onward because all the further calculations are done off-shell)
\begin{align}
\sqrt{-g}J^i_{ADT}=\d(\sqrt{-g}E^{ij}\xi_j)-\frac{1}{2}\sqrt{-g}\xi^iE^{jk}h_{jk}~. \label{JADTOFFG}
\end{align}
The above relation follows from the fact that $\d\xi^i=0$ and $\d\phi=0$ as we consider only the change due to $g_{ab}\rightarrow g_{ab}+h_{ab}$.
By varying the Noether current in Jordan frame {\it i.e} eq.(\ref{J'A}), under the change in the metric tensor, we obtain
\begin{align}
\d(\sqrt{-g}J'^i)=2\d(\sqrt{-g}E^{ij}\xi_j)-\sqrt{-g}\xi^iE^{jk}h_{jk}
\no 
\\
+\sqrt{-g}\xi^i\na_b[\T'^b(q,\d q)]-\d[\sqrt{-g}\T'^i(q,\lie q)]~.
\end{align}
Using the equation \eqref{JADTOFFG}, the above equation reduces to:
\begin{align}
\d(\sqrt{-g}J'^i)=2\sqrt{-g}J^i_{ADT}+\sqrt{-g}\xi^i\na_b[\T'^b(q,\d q)]
\no 
\\
-\d[\sqrt{-g}\T'^i(q,\lie q)]~.\label{VARJI}
\end{align}
 As $\xi^a$ is the Killing vector, in the above expression, therefore we use $\d[\sqrt{-g}\T'i(q,\lie q)]=\lie[\sqrt{-g}\T'^i(q,\d q)]$ ( which follows from the fact that $\o^a=0$ in \eqref{O1}). Using this relation in the above equation, it is straightforward to obtain 
\begin{align}
\sqrt{-g}J^{ij}_{ADT}=\frac{1}{2}\d(\sqrt{-g}J'^{ij})-\sqrt{-g}\xi^{[i}\T'^{j]}(q,\d q)~. \label{NOADT}
\end{align}
In the Einstein frame, by following the similar steps one finally obtains, 
\begin{align}
\sqrt{-\t g}\t J^{ij}_{ADT}=\frac{1}{2}\d(\sqrt{-\t g}\t J^{ij})-\sqrt{-\t g}\t \xi^{[i}\t \T^{j]}(\t q,\d \t q)~. \label{NOADT'}
\end{align}
The above equations \eqref{NOADT} and \eqref{NOADT'} show the explicit connections between the ADT and the Noether conserved quantities in the two frames.

Now we intend to show that how the ADT potentials are conformally connected in the two frames. Using \eqref{JABJAB'} and \eqref{THTH'} in the equations \eqref{NOADT} and \eqref{NOADT'}, it can be easily shown that,
\begin{align}
\t J^{ij}_{ADT}=\frac{J^{ij}_{ADT}}{\phi^2}~.
\end{align}
 Thus our result is implying that the ADT potentials in the two frames are conformally connected to each other in the same manner as we obtain in the case of Noether potentials. Thus, the conserved ADT charges are invariant in the two frames. Such a prediction was given earlier in \cite{Deser:2006gt}.
 
Let us now conclude this section with the following comments. In Komar's method \cite{Komar:1958wp} of defining the mass and angular momentum at the asymptotic infinity by using the conserved Noether current, there appears an anomalous factor of $2$ \cite{Katz}. This anomaly can be tackled by the background subtraction method as described in \cite{Katz}. Later, Wald provided an elegant solution in this context by considering the variation of the Noether current (we implemented the similar analysis in the section \eqref{WALD}) and defined the mass and the angular momentum in terms of the integrals containing the Noether current along with the correction term (as in eq. \eqref{DELH2}), which resolves the anomalous $2$.
 
Apart from the Wald's formalism, the first law can be established from the conserved ADT currents as well \cite{Hyun:2017nkb}. One can see from \eqref{NOADT} and \eqref{NOADT'}, the ADT potential consists of the Noether potential along with the same extra correction term which appears in the Wald's formalism (as in eq. \eqref{DELH2}). The similar idea has also been successfully adopted in other spacetime solutions like Lifshitz black holes \cite{Ayon-Beato:2015jga}. Thus we emphasize that the both way of establishing the first law are equivalent and, therefore these two methods can be implemented alternatively according to one's convenience. For the reasons stated above, we do not include the explicit calculation of establishing the first law using the ADT formalism.
%%%%%%%%%%%%%%%%%%%%%%%%%%%%%%%%%%%%%%%%%%%%%%%%%%%%%%%%%%%%%%%%%%%%%%%%%%%%%%%
\section{Conclusions} \label{CON}
It is widely known that the correspondence between the thermodynamic quantities and  the spacetime geometry is not confined only to the Einstein's theory of general relativity. Moreover, the recent experimental developments are strongly suggesting us that, in order to get a complete understanding of general relativity, one should move on to the studies of modified theories of gravity and their implications in gravitational and cosmological phenomena. 
The scalar-tensor theory is one of the most popular among the alternative theories of gravity but the underlying thermodynamic description of this theory is not yet properly developed. 
Some of the ambiguities are as follows: We have mentioned earlier that the scalar tensor theory can be analysed in the Jordan frame and as well as in the Einstein frame. Untill now there is no exact covariant expression of energy which can fit across all the thermodynamic aspects of the theory in these two frames.
Moreover in this connection, there are several conflicts between the physicists regarding the description of thermodynamic quantities and their invariance in these two frames.

 In literature, the authors of \cite{Koga:1998un} have shown that in the background of an asymptotically flat spacetime, the thermodynamic quantities are conformally equivalent in these two frames in the scalar tensor theory. In our previous work \cite{Bhattacharya:2017pqc}, we have shown that the thermodynamic quantities must be conformally invariant but, in that framework we could not provide the exact covariant expression of the energy.  However, there is another standard mechanism in literature called ADT formalism, which also can be used as a tool to understand the thermodynamic properties of spacetime geometry.
 We emphasize on this point that the critical study of thermodynamic properties using the ADT formalism has not been analysed yet in the background of the scalar-tensor theory.

In this work, we intend to cast light on the above issues and provide satisfactory answers to all these incongruities in this theory. We start from the basic action level of the scalar-tensor theory and show that the usual Lagrangian in the two frames differ by a total derivative term due to the conformal transformation. It is common in the study of scalar-tensor theory, that most of the authors does not carefully mention that the two Lagrangians in Jordan frame and the Einstein frame are equivalent only up to a total derivative term. Although a total derivative term does not contribute to the dynamics of the system, but one must contemplate deeply before one injudiciously neglect that term in this theory while studying the thermodynamic aspects. In our present work, we show that this surface term actually plays the crucial role to obtain the conformal equivalence of the thermodynamic quantities without imposing any assumptions and boundary conditions.

In this work, the study of thermodynamic properties of spacetime geometry is based on the concept of conserved currents as obtained from the two different approaches such as the Noether approach and the ADT approach. All the conserved quantities are off-shell, which can be used for a generic null surface and can play a significant role in the context of the emergent gravity paradigm.
At first,  we obtain the off-shell Noether current and potential in both the frames and,  following the Wald's formalism, we identify the thermodynamic quantities from the conserved Noether current. Later, we show that the identified thermodynamic quantities fit nicely in the first law and the second law of black hole mechanics. Subsequently we obtain an important result in the background of our theoretical framework that the thermodynamic parameters are conformally invariant in these two frames, if one consider the $\square\phi$ term in the Lagrangian. Hence, at this stage, we comment that to examine the conformal invariance of the thermodynamic quantities in the two frames in the background of  the scalar-tensor theory, one must not disregard the contributions from the surface term. We also emphasize that following our procedure of {\i.e} the inclusion of $\square\phi$ term, one can avoid the use of any boundary condition and assumption regarding the nature of spacetime.
Observing the above conclusions in Noether prescription, we are keen to verify our results using the ADT formalism in both the frames of this theory. Therefore following the ADT mechanism, we obtain the conserved ADT current and the corresponding ADT potential in both the frames. Thereafter, we establish the connection of the ADT current and potential to the Noether counterparts. Moreover, we discuss the connection of the off-shell ADT currents with the off-shell Noether current and Wald's formalism. Our results strongly support that, implementing both of these standard formalism we find that the thermodynamic descriptions and the thermodynamic quantities are invariant in the two frames in the background of the scalar-tensor theory. Our results suggest that these two approaches of finding conserved quantities and describing the first law of black hole thermodynamics are basically equivalent to each other. We hope this work will be a significant one in the thermodynamic description of the scalar-tensor theory.

Finally, we mention that in usual thermodynamics there are intensive quantities (such as temperature and pressure etc.) which do not change
by conformal scalings, while there are extensive quantities (like energy) which do change under scaling. Our present situation is in contradiction with this usual understanding. This issue can be understood in the following way. In black hole thermodynamics, we cannot categorize the extensive and the intensive variables like the usual thermodynamic cases. For example, the entropy is an extensive variable and also a function of all other extensive parameters in the usual thermodynamics. But, in black hole mechanics it is not an extensive variable as it is proportional to the area of the black hole horizon. If the two black holes are combined together, then Bekenstein-Hawking area expression implies that the entropy of the combined black hole is greater than the sum of the entropy of the individual black holes. Moreover, the temperature and pressure in the usual thermodynamic case are intensive thermodynamic entities. But, in black hole thermodynamics, those two quantities are scale dependent (for instance, in the case of Schwarzschild black hole the Hawking temperature is inverse of mass of the black hole). The principal of equivalence implies that the temperature is red-shifted or blue-shifted in the same manner as of the frequency of the photons. Apart from these obvious differences with the usual thermodynamics, there are a few other facts (e.g. specific heat of Schwarzschild black hole is negative) which clearly indicates that one cannot classify the black hole thermodynamic entities as the extensive or the intensive ones.  Therefore, the usual scaling argument cannot be applied here. 

%%%%%%%%%%%%%%%%%%%%%%%%%%%%%%%%%%%%%%%%%%%%%%%%%%%

\vskip 4mm
\appendix
\section{Derivation of the Eq. \eqref{JAB'}} \label{APPEN1}
The expression of the of $\T'^a(q, \lie q)$ can be obtained from \eqref{EXACTEXPJOR}. Let us calculate term by term. At first we obtain
\begin{eqnarray}
&&2(\nabla_b\phi)P^{iabd}\pounds_{\xi} g_{id}
\nonumber
\\
&&=(\nabla^d\phi)[\nabla^a\xi_d+\nabla_d\xi^a]-2(\nabla^a\phi)(\nabla_i\xi^i); \label{A2}
\end{eqnarray}
and
\begin{eqnarray}
&& \pounds_{\xi}v^a=2P^{ibad}\nabla_b\pounds_{\xi}g_{id}=2P^{iabd}\nabla_b\pounds_{\xi}g_{id}
\nonumber
\\
&&=\nabla_b\nabla^a\xi^b+\nabla_b\nabla^b\xi^a-2\nabla^a\nabla_b\xi^b~.  \label{A3}
\end{eqnarray}
Hence we obtain,
\begin{eqnarray}
&&2(\nabla_b\phi)P^{iabd}\pounds_{\xi} g_{id}-\phi\pounds_{\xi}v^a
\nonumber
\\
&&=(\nabla_b\phi)(\nabla^a\xi^b)+\phi\nabla_b\nabla^a\xi^b-\phi\square\xi^a 
\nonumber
\\
&&+(\nabla_b\phi)(\nabla^b\xi^a)-2(\nabla^a\phi)(\nabla_b\xi^b)-2\phi g^{ac}R_{kc}\xi^{k}, 
\no 
\\
\label{A4}
\end{eqnarray}
Here, we have used $[\nabla_b\nabla_d]\xi_i=R_{ijbd}\xi^j$ to obtain the last term. Now,
\begin{align}
(\nabla_b\phi)(\nabla^a\xi^b)+\phi\nabla_b\nabla^a\xi^b-\phi\square\xi^a=\nabla_b[\phi(\nabla^a\xi^b-\nabla^b\xi^a)]
\no 
\\
+(\nabla_b\phi)(\nabla^b\xi^a)~, \label{A5}
\end{align}
and
\begin{align}
2(\nabla_b\phi)(\nabla^b\xi^a)-2(\nabla^a\phi)(\nabla_b\xi^b)=2\nabla_b[\xi^a(\nabla^b\phi)-\xi^b(\nabla^a\phi)]
\no 
\\
+2\xi^b\nabla_b\nabla^a\phi-2\xi^a\square\phi~. \label{A6}
\end{align}
Substituting the values of \eqref{A5} and \eqref{A6} in \eqref{A4}, we finally obtain 
\begin{eqnarray}
&&2(\nabla_b\phi)P^{iabd}\pounds_{\xi} g_{id}-\phi\pounds_{\xi}v^a
\nonumber
\\
&&=\nabla_b[\phi(\nabla^a\xi^b-\nabla^b\xi^a)+2\xi^a(\nabla^b\phi)-2\xi^b(\nabla^a\phi)]
\nonumber
\\
&&+2\xi^b\nabla_b\nabla^a\phi-2\xi^a\square\phi-2\phi g^{ac}R_{kc}\xi^k~. \label{A7}
\end{eqnarray}
Moreover, a straightforward calculations give (the extra contributions from the $\square\phi$ term)
\begin{align}
\frac{3}{2}g^{ij}\lie g_{ij}\p^a\phi-3g^{ia}\p^b\phi\lie g_{ib}+3\p^a(\lie\phi)
\no 
\\
=3\na_b[\xi^b\na^a\phi-\xi^a\na^b\phi]+3\xi^a\square\phi~.
\end{align}
 
Thus we finally obtain
\begin{align}
\T'^a(q,\lie q)=\frac{1}{16\pi}\Big[-\na_b\big[\nabla^a(\phi\xi^b)-\nabla^b(\phi\xi^a)\big]
\no 
\\
-\frac{2\o}{\phi}(\na^a\phi)\xi^b\na_b\phi-2\xi^b\na_b\na^a\phi-\xi^a\square\phi+2\phi g^{ac}R_{kc}\xi^k\Big]~. \label{AA}
\end{align}
Using \eqref{AA} and $L'$ from \eqref{L'} in \eqref{J'A} yields
\begin{eqnarray}
&&J'^a=\frac{1}{16\pi}\Big[\big[\nabla^a(\phi\xi^b)-\nabla^b(\phi\xi^a)\big]+\Big\{(\phi R
\nonumber
\\
&& -\frac{\omega (\phi)}{\phi}g^{ab}\nabla_a\phi \nabla_b\phi-V(\phi))\xi^a+2\frac{\omega}{\phi}(\nabla^a\phi)\xi^b (\nabla_b\phi)
\nonumber
\\
&&+2\xi^b\nabla_b\nabla^a\phi-2\xi^a\square\phi-2\phi g^{ac}R_{kc}\xi^k\Big\}+2E^{ab}\xi_b
\Big].
\no 
\\
 \label{A8}
\end{eqnarray}
One can identify the second-bracketed term as a whole as $-2E^{ab}\xi_b$ (see the expression of $E^{ab}$ from \eqref{EXACTEXPJOR}) and, hence, the expression of $J'^a$ is given by a total derivative of anti-symmetric Noether potential, the expression of which has been given in \eqref{JAB'}.
%%%%%%%%%%%%%%%%%%%%%%%%%%%%%%%%%%%%%%%%%%%%%%%%%%%%%%%%%%%%%%%%%%%%%
\section{Derivation of the Eqs. \eqref{JEINT} and \eqref{JABEIN}} \label{APPEN3}
The exact expression of $\t\T^a (\t q,\tlie \t q)$ can be obtained from \eqref{EXACTEXEIN}. Straightforwardly, one can obtain 
\begin{eqnarray}
&&\pounds_{\xi}\tilde{v}^a=\tilde{\nabla}_b\tilde{\nabla}^a\tilde{\xi}^b+\tilde{\nabla}_b\tilde{\nabla}^b\tilde{\xi}^a-2\tilde{\nabla}^a\tilde{\nabla}_b\tilde{\xi}^b
\nonumber
\\
&&=\tilde{\nabla}_b\tilde{\nabla}^b\tilde{\xi}^a-\tilde{\nabla}_b\tilde{\nabla}^a\tilde{\xi}^b+2\tilde{g}^{ac}\tilde{R}_{kc}\tilde{\xi}^k~. \label{C2}
\end{eqnarray}
Thus, from \eqref{JEINT} one can obtain
\begin{eqnarray}
&&\tilde{J}^a=\Big\{\Big(\frac{\tilde{R}}{16\pi}-\frac{1}{2}\tilde{g}^{ij}\tilde{\nabla}_i\tilde{\phi}\tilde{\nabla}_j\tilde{\phi}-U(\tilde{\phi})\Big)\tilde{\xi}^a+(\tilde{\nabla}^a\tilde{\phi})\tilde{\xi}^b(\tilde{\nabla}_b\tilde{\phi})
\nonumber
\\
&&-\frac{2}{16\pi}\tilde{g}^{ac}\tilde{R}_{kc}\tilde{\xi}^k\Big\}+\frac{1}{16\pi}\tilde{\nabla}_b[\tilde{\nabla}^a\tilde{\xi}^b-\tilde{\nabla}^b\tilde{\xi}^a]+2\t E^{ab}\t\xi_b~.
\end{eqnarray}
The second-bracketed terms, as a whole, contribute as $-2\t E^{ab}\t\xi_b$ and, therefore, $\tilde{J}^a$ can be written as a total derivative term as $\tilde{J}^a=\t\na_b\t J^{ab}$. Thus the expression of $\t J^{ab}$ will be of the form given in \eqref{JABEIN}.
%%%%%%%%%%%%%%%%%%%%%%%%%%%%%%%%%%%%%%%%%%%%%%%%%%%%%%%%%%%%%%%%%%%%%
\section{Derivation of the Eqs. \eqref{JABJAB'} and \eqref{THTH'}}\label{APPEN4}
Proving \eqref{JABJAB'} is pretty straightforward. 
\begin{align}
\t J^{ab}=\t g^{ai}\t g^{bj}\t J_{ij}=\t g^{ai}\t g^{bj}(\p_a\t\xi_b-\p_b\t\xi_a)
\no 
\\
=\frac{g^{ai} g^{bj}}{\phi^2}\Big[\p_a(\phi\xi_b)-\p_b(\phi\xi_a)\Big]\ \ 
\no 
\\
=\frac{1}{\phi^2}\Big[\na^a(\phi\xi^b)-\na^b(\phi\xi^a)\Big]~.
\end{align}
Thus, equation \eqref{JABJAB'} is obtained.

The expression of $\t\T^a$ is given in \eqref{EXACTEXEIN}. Now, $\t\na_b(\d \t g_{id})=(\p_b\phi)\d g_{id}+\phi\t\na_b(g_{id})-\frac{g_{id}}{\phi}(\p_b\phi)\d\phi+g_{id}\p_b(\d\phi)$. Then using $\t\Gamma^a_{bc}=\Gamma^a_{bc}+\frac{1}{2\phi}(\d^a_b\p_c\phi+\d^a_c\p_b\phi-g_{bc}\p^a\phi)$ in $\t\na_b(g_{id})$, it requires a few steps to obtain \eqref{THTH'}.
%%%%%%%%%%%%%%%%%%%%%%%%%%%%%%%%%%%%%%%%%%%%%%%%%%%%%%%%%%%%%%%%%%%%%%
\section{Derivation of the Eq. \eqref{JIJADT}} \label{APPEN5}
For $g_{ab}\rightarrow g_{ab}+h_{ab}$, the expression of $\d G^{ij}\xi_{j}$ is given as \cite{Bouchareb:2007yx}
\begin{align}
(\d G^{ij})\xi_{j}=\na_j F^{ij}-G^{ik}h_{kj}\xi^j+\frac{1}{2}\xi^i G^{jk}h_{jk}-\frac{1}{2}\xi^jG^i_jh~, \label{E1}
\end{align}
where, $\d G^{ij}$ denotes the linearization of the Einstein tensor. Remember, here $\xi^a$ is a Killing vector and
\begin{align}
F^{ij}=\frac{1}{2}\Big[\xi^j\na_kh^{ki}-\xi^i\na_kh^{kj}+\xi_k\na^ih^{kj}-\xi_k\na^jh^{ki}
\no 
\\
+\xi^i(\na^jh)-\xi^j(\na^ih)+h^{kj}\na_k\xi^i-h^{ki}\na_k\xi^j+h\na^{[i}\xi^{j]} \Big]~. \label{E2}
\end{align}
Now, in this frame the expression of $E^{ab}$ has been given in \eqref{EXACTEXPJOR} . For  $g_{ab}\rightarrow g_{ab}+h_{ab}$
\begin{align}
16\pi(\d E^{ij})\xi_j=\phi[(\d G^{ij})\xi_j]-\frac{\o}{2\phi}h^{ij}g^{ab}(\p_a\phi)(\p_b\phi)\xi_j
\no 
\\
-\frac{\o}{2\phi}\xi^ih^{ab}(\p_a\phi)(\p_b\phi)+\frac{\o}{\phi}h^{jb}g^{ai}(\p_a\phi)(\p_b\phi)\xi_j-\frac{V}{2}h^{ij}\xi_j
\no 
\\
+h^{ia}\xi^b(\na_a\na_b\phi)+g^{ia}h^{jb}\xi_j(\na_a\na_b\phi)-h^{ij}g^{ab}\xi_{j}(\na_a\na_b\phi)
\no 
\\
-\xi^ih^{ab}(\na_a\na_b\phi)-g^{ia}\xi^b\d(\na_a\na_b\phi)+\xi^ig^{ab}\d(\na_a\na_b\phi)~. \label{E3}
\end{align}
Now, we express $16\pi E^{ij}=\phi G^{ij}+\bar{E}^{ij}$ where
\begin{align}
 \bar{E}^{ij}=\frac{\o}{2\phi}g^{ij}g^{ab}(\p_a\phi)(\p_b\phi)-\frac{\o}{\phi}g^{ia}g^{jb}(\p_a\phi)(\p_b\phi)
 \no 
 \\
+\frac{V}{2}g^{ij} -g^{ia}g^{jb}\na_a\na_b\phi+g^{ij}g^{ab}\na_a\na_b\phi~.
\end{align}
Then, using \eqref{E1} and \eqref{E3} we obtain
\begin{eqnarray}
&&16\pi(\d E^{ij})\xi_j=\phi\na_j F^{ij}-16\pi E^{ik}h_{kj}\xi^j+\frac{16\pi}{2}\xi^i E^{jk}h_{jk}
\no 
\\
&&-\frac{16\pi}{2}\xi^jE^i_jh+\bar{E}^{ik}h_{kj}\xi^j-\frac{1}{2}\xi^i \bar{E}^{jk}h_{jk}+\frac{1}{2}\xi^j\bar{E}^i_jh
\no 
\\
&&-\frac{\o}{2\phi}h^{ij}g^{ab}(\p_a\phi)(\p_b\phi)\xi_j-\frac{\o}{2\phi}\xi^ih^{ab}(\p_a\phi)(\p_b\phi)
\no 
\\
&&+\frac{\o}{\phi}h^{jb}g^{ai}(\p_a\phi)(\p_b\phi)\xi_j-\frac{V}{2}h^{ij}\xi_j+h^{ia}\xi^b(\na_a\na_b\phi)
\no 
\\
&&+g^{ia}h^{jb}\xi_j(\na_a\na_b\phi)-h^{ij}g^{ab}\xi_{j}(\na_a\na_b\phi)-\xi^ih^{ab}(\na_a\na_b\phi)
\no 
\\
&&-g^{ia}\xi^b\d(\na_a\na_b\phi)+\xi^ig^{ab}\d(\na_a\na_b\phi)~. \label{E4}
\end{eqnarray}
Now,
\begin{align}
\bar{E}^{ik}h_{kj}\xi^j-\frac{1}{2}\xi^i \bar{E}^{jk}h_{jk}+\frac{1}{2}\xi^j\bar{E}^i_jh=\frac{\o}{2\phi}h^{ij}g^{ab}(\p_a\phi)(\p_b\phi)\xi_j
\no 
\\
+\frac{V}{2}h^{ij}\xi_j-g^{ia}h^{jb}\xi_j(\na_a\na_b\phi)+h^{ij}g^{ab}\xi_{j}(\na_a\na_b\phi)
\no 
\\
-\frac{\o}{\phi}h^{jb}g^{ai}(\p_a\phi)(\p_b\phi)\xi_j+\frac{\o}{2\phi}\xi^ih^{ab}(\p_a\phi)(\p_b\phi)
\no 
\\
+\frac{1}{2}\xi^ih^{ab}(\na_a\na_b\phi)-\frac{1}{2}g^{ia}\xi^bh(\na_a\na_b\phi)~. \label{E5}
\end{align}
Substituting \eqref{E5} in \eqref{E4}, we obtain
\begin{align}
16\pi(\d E^{ij})\xi_j=\na_j (\phi F^{ij})-F^{ij}(\p_j\phi)-16\pi E^{ik}h_{kj}\xi^j
\no 
\\
+\frac{16\pi}{2}\xi^i E^{jk}h_{jk}-\frac{16\pi}{2}\xi^jE^i_jh+h^{ia}\xi^b(\na_a\na_b\phi)
\no 
\\
-\frac{1}{2}\xi^ih^{ab}(\na_a\na_b\phi)-g^{ia}\xi^b\d(\na_a\na_b\phi)
\no 
\\
+\xi^ig^{ab}\d(\na_a\na_b\phi)-\frac{1}{2}g^{ia}\xi^bh(\na_a\na_b\phi)~. \label{E6}
\end{align}
Now,
\begin{align}
\d(\na_b\na_a\phi)=-\d \Gamma^i_{ab}(\p_i\phi)
\no 
\\
=-\frac{1}{2}\Big[\na_a h^i_b+\na_b h^i_a-\na^ih_{ab} \Big](\p_i\phi)~. \label{E7}
\end{align}
Using the above relation \eqref{E7} with \eqref{E2}, one obtains
\begin{align}
-F^{ij}(\p_j\phi)-g^{ia}\xi^b\d(\na_a\na_b\phi)+\xi^ig^{ab}\d(\na_a\na_b\phi)
\no 
\\
=\frac{1}{2}(\p_j\phi)\Big[-\xi^i\na_kh^{jk}-h^{jk}(\na_k\xi^i)+h^{ik}(\na_k\xi^j)
\no 
\\
-h\na^{[i}\xi^{j]}+\xi^k\na_kh^{ij} \Big]
\no 
\\
=\frac{1}{2}\na_j\Big[(\p_k\phi)\Big(\xi^jh^{ik}-\xi^ih^{jk}\Big)\Big]+\frac{1}{2}h^{jk}\xi^i\na_k\na_j\phi
\no 
\\
+\frac{1}{2}h^{ik}(\na_k\xi^j)(\na_j\phi)-\frac{1}{2} h(\na_j\phi)(\na^i\xi^j)-\frac{1}{2}h^{ij}\xi^k\na_k\na_j\phi~. \label{E8}
\end{align}
Substituting the above relation of \eqref{E8} in \eqref{E6}, we obtain
\begin{eqnarray}
&&16\pi(\d E^{ij})\xi_j=\na_j (\phi F^{ij})+\frac{1}{2}\na_j\Big[(\p_k\phi)\Big(\xi^jh^{ik}-\xi^ih^{jk}\Big)\Big]
\no 
\\
&&-16\pi E^{ik}h_{kj}\xi^j+\frac{16\pi}{2}\xi^i E^{jk}h_{jk}-\frac{16\pi}{2}\xi^jE^i_jh
\no 
\\
&&+\frac{1}{2}\Big\{ h^{ia}\xi^b(\na_a\na_b\phi)-g^{ia}\xi^bh(\na_a\na_b\phi)+h^{ik}(\na_k\xi^j)(\na_j\phi)
\no 
\\
&&-h(\na_j\phi)(\na^i\xi^j)\Big\}~. \label{E9}
\end{eqnarray}
Using the property of the Killing vector, the terms inside the curly bracket vanish and, one obtains
\begin{align}
(\d E^{ij})\xi_j=\na_j J^{ij}_{ADT}-E^{ik}h_{kj}\xi^j
\no 
\\
+\frac{1}{2}\xi^i E^{jk}h_{jk}-\frac{1}{2}\xi^jE^i_jh~, \label{E10}
\end{align}
where, the final expression of $J^{ij}_{ADT}$ is given in \eqref{JIJADT}.
%%%%%%%%%%%%%%%%%%%%%%%%%%%%%%%%%%%%%%%%%%%%%%%%%%%%%%%%%%%%%%%%%%%%%
\section{Derivation of the Eq. \eqref{JIJADTTIL}} \label{APPEN6}
To prove \eqref{JIJADTTIL}, we shall follow the same procedure as in the Jordan frame. Here, let us take $\t E^{ij}=\frac{\t G^{ij}}{16\pi}+\bar{\t E}^{ij}$ with
\begin{align}
\bar{\t E}^{ij}=-\frac{1}{2}\t g^{ai}\t g^{bj}(\p_a\t\phi)(\p_b\t\phi)+\frac{1}{4}\t g^{ij}\t g^{ab}(\p_a\t\phi)(\p_b\t\phi)+\frac{1}{2}\t g^{ij}U~.
\end{align} 
Therefore,
\begin{eqnarray}
&&(\d\t E^{ij})\t\xi_j=\frac{1}{16\pi}(\d\t G^{ij})\t\xi_j+(\d \bar{\t E}^{ij})\t\xi_j
\no 
\\
&&=\frac{1}{16\pi}\t\na_j\t F^{ij}-\t E^{ik}\t h_{kj}\t \xi^j+\frac{1}{2}\t\xi^i\t E^{jk}\t h_{jk}
\no 
\\
&&-\frac{1}{2}\t\xi^j\t E^i_j\t h+\Big\{ \bar{\t E}^{ik}\t h_{kj}\t \xi^j-\frac{1}{2}\t\xi^i \bar{\t E}^{jk}\t h_{jk}+\frac{1}{2}\t\xi^j \bar{\t E}^i_j\t h+(\d \bar{\t E}^{ij})\t\xi_j\Big\}~.
\no 
\\ \label{F1}
\end{eqnarray}
where, the expression of $\t F^{ij}$ is similar to the expression given in \eqref{E2} (only with tilde overhead).
Detail calculations show that the terms inside the curly brackets in \eqref{F1} vanish and, one finally obtains
\begin{align}
(\d\t E^{ij})\t\xi_j=\t\na_j\t J^{ij}_{ADT}-\t E^{ik}\t h_{kj}\t \xi^j+\frac{1}{2}\t\xi^i\t E^{jk}\t h_{jk}-\frac{1}{2}\t\xi^j\t E^i_j\t h~,
\end{align}
where, the final expression of $\t J^{ij}_{ADT}$ is given in \eqref{JIJADTTIL}.
%%%%%%%%%%%%%%%%%%%%%%%%%%%%%%%%%%%%%%%%%%%%%%%%%%%%%%%%%%% 

\end{document}